\newlength{\extralineskip}
\documentclass[12pt]{article}
\usepackage{epsfig}

\begin{document}
\begin{titlepage}
\begin{flushright}
          \begin{minipage}[t]{12em}
          \large UAB--FT--448\\
                 July 1997
          \end{minipage}
\end{flushright}
\vspace{\fill}

\vspace{\fill}

\begin{center}
\baselineskip=2.5em

{\large \bf Long range forces induced by neutrinos at finite temperature}
\end{center}

\vspace{\fill}

\begin{center}
{\bf F. Ferrer, J.A. Grifols, and M. Nowakowski}\\
\vspace{0.4cm}
     {\em Grup de F\'\i sica Te\`orica and Institut de F\'\i sica
     d'Altes Energies\\
     Universitat Aut\`onoma de Barcelona\\
     08193 Bellaterra, Barcelona, Spain}
\end{center}
\vspace{\fill}

\begin{center}
\large Abstract
\end{center}
\begin{center}
\begin{minipage}[t]{36em}
We revisit and extend  previous work on neutrino mediated long range forces in a
background at finite temperature. For Dirac neutrinos, we correct existing
results. We also give new results concerning spin-independent as well as
spin-dependent long range forces associated to Majorana neutrinos. An
interesting outcome of the investigation is that, for both types of
neutrinos whether massless or not, the effect of the relic neutrino heat
bath is to convert those forces into attractive ones in the supra-millimeter
scale while they stay repulsive within the sub-millimeter
scale.   
\end{minipage}
\end{center}

\vspace{\fill}

\end{titlepage}

\clearpage

\addtolength{\baselineskip}{\extralineskip}

Neutrinos mediate long-range forces between macroscopic bodies
\cite{fs}, \cite{hs}, \cite{fey}, \cite{gri1}, \cite{fisch}. 
Indeed double
neutrino exchange among matter fermions generates spin-independent forces
that extend coherently over macroscopic distances. The effect, however, is
extremely weak, much too tiny to be experimentally detected with present
day technology. 
Compared to
their gravitational pull, the force between two nucleons 1 cm apart is about
$10^{-28}$ times weaker. Not only their coupling strength
is very small but also their decay with distance is fast. Indeed the potential drops as
$r^{-5}$ so that the effects die off correspondingly. Phenomenological surveys on forces
with this particular distance behaviour have been conducted in the literature (see
e.g.\cite{gri2}) over the whole span of distances from astronomical down
to the micron scale. If at all, this forces will induce physical effects in
the sub-millimeter (but macroscopic) end of the distance scale. Perhaps an exception to
this is the case of a system with high density of matter such as the core of a neutron
star where collective effects may show up \cite{fisch}, \cite{aba}. 

 In a neutrino populated medium, such as the cosmic neutrino
background or the hot core of a supernova, the helicity flip produced by single neutrino
exchange can be balanced  by the neutrinos in the medium and, as a consequence, a
spin-independent interaction takes place that leads to a coherent effect over many
particles in bulk matter.  

The neutrino long-range forces in the presence of a neutrino thermal bath
have been explored in reference \cite{hp} in the Dirac neutrino case. The long
range forces mediated by Majorana neutrinos, on the other hand, have been
studied only in the zero temperature case \cite{gri1}. 
Here we wish to extend the 
nonzero temperature results to the Majorana case. Because the distinction
between Dirac and Majorana neutrinos is superfluous for massless neutrinos,
we shall consider the general $m\not= 0$ case. 

We shall adopt the notation in \cite{hp} and write,

\begin{equation}\label{potential}
V(r)=-\int {d^{3}{\bf Q} \over(2\pi)^{3}} \exp (i{\bf Q}\cdot {\bf r}) T({\bf Q})
\end{equation}
where $T({\bf Q})$ 
is the nucleon-nucleon elastic scattering amplitude (figure 1) in the static
limit, i.e. momentum transfer $Q\simeq(0,{\bf Q})$, where matter is supposed to be at
rest in the microwave background radiation (MWBR) frame. It can be cast in the form 

\begin{equation}\label{amplitude} 
T(Q)=-2iG_{F}^2 (g_{V},-2g_{A}{\bf S})^{\mu}(g'_{V},-2g'_{A}
{\bf S}')^{\nu}I_{\mu \nu}
\end{equation}
with 
\begin{equation}
I_{\mu \nu}=\int {d^{4}k \over(2\pi)^4}
Tr[\gamma_{\mu}OiS_{T}(k)\gamma_{\nu}OiS_{T}(k-Q)].
\end{equation}

\begin{figure}[bht]
\begin{center}
\epsfig{file=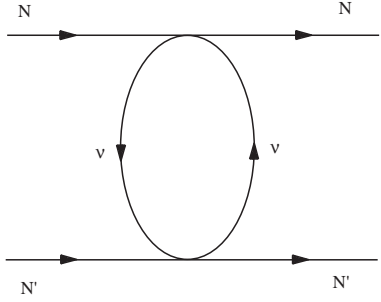,width=5cm,height=4cm}
\end{center}
\caption{Fig. 1. {\it Lowest order Feynman diagram for two neutrino exchange
    in the four fermion effective theory.}}
\end{figure}

The operator $O$ is the
left-handed projector ${1\over 2}(1-\gamma_{5})$ for Dirac neutrinos and
${\sqrt2\over 2}\gamma_{5}$ for Majorana neutrinos. The temperature dependent
propagator $S_{T}$ has the explicit form

\begin{equation}\label{propagator}
S_{T}(k)=(\rlap / k +m)\left[(k^2-m^2+i\epsilon)^{-1}+2\pi i\delta(k^2-m^2)(\theta
(k^0)n_{+}+\theta(-k^0)n_{-})\right]
\end{equation}
where $n_{+}$ and $n_{-}$ are Fermi-Dirac distribution functions for
particle and antiparticle, respectively. As discussed in \cite{hp}, figure 1
evaluated with this propagator taken together with the usual Feynman rules is
sufficient to calculate the potential. In equation (2),
$g_{V,A}$ are composition-dependent weak vector and axial-vector couplings. We focus
first on the spin-independent potential, that is the $g_{V}g_{V}'$ component of equation
(2). 

Use of the first piece in equation (4) gives the zero temperature vacuum
results \cite{gri1}, \cite{fisch},
\begin{equation}\label{dirac}
V_{Dirac}(r)={G_{F}^2m^3g_{V}g_{V}'\over 4\pi^3r^2}K_{3}(2mr)
\end{equation}
and
\begin{equation}\label{majorana}
V_{Majorana}(r)={G_{F}^2m^2g_{V}g_{V}'\over 2\pi^3r^3}K_{2}(2mr)
\end{equation}
in terms of the modified Bessel functions $K_{2,3}$.

At very large distances ($mr\gg 1$), i.e. much larger than the Compton
wavelength of the neutrino, these potentials exhibit the asymptotic behaviour

\begin{equation}\label{dirac ld}
V_{Dirac}(r)\simeq {G_{F}^2g_{V}g_{V}'\over 8}\left({m\over \pi r} \right)^{5/2}e^{-2mr}
\end{equation}
and
\begin{equation}\label{majorana ld}
V_{Majorana}(r)\simeq {G_{F}^2g_{V}g_{V}'\over4}\left({m^3\over
\pi^5r^7}\right)^{1/2}e^{-2mr}.
\end{equation}
Of course, both potentials (equations (5) and (6)) coincide when $m=0$. They
give the well known Feinberg and Sucher result:
\begin{equation}\label{feinbergsucher}
V(r)={G_{F}^2g_{V}g_{V}'\over 4\pi^3r^5}.
\end{equation}
In a neutrino background, a contribution to the long range force can arise
because a neutrino in the thermal bath may be excited and de-excited back to its
original state in the course of the double scattering process. This effect is
described by the crossed terms contained in $I_{\mu\nu}$ that involve the
thermal piece of one neutrino propagator along with the vacuum piece of the
other neutrino propagator. This thermal component of the tensor $I_{\mu\nu}$
can be written as

\begin{eqnarray}
I^{\mu \nu}_{T,D}&=&-\pi
i\int{d^{4}k\over(2\pi)^4}\delta(k^2-m^2)[\theta(k^0)n_{+}+\theta(-k^0)n_{-}]
\nonumber \\
&&\times \left[{Tr\left[\gamma^{\mu}
(\rlap /k+\rlap /Q)\gamma^{\nu} \rlap / k\right]\over (k+Q)^2-m^2+i\epsilon}+{Tr\left[\gamma^{\mu}\rlap / k\gamma^{\nu} (\rlap /k-\rlap /Q)\right]\over 
(k-Q)^2-m^2+i\epsilon}\right]
\end{eqnarray}
in the Dirac case, and
\begin{eqnarray}
I^{\mu \nu}_{T,M}&=&
-\pi
i\int{d^{4}k\over(2\pi)^4}\delta(k^2-m^2)\: n \nonumber \\
&&\times \left[{Tr\left[\gamma^{\mu}
(\rlap /k+\rlap /Q+m)\gamma^{\nu} (\rlap /k-m)\right]\over (k+Q)^2-m^2+i\epsilon}+{Tr\left[\gamma^{\mu} (\rlap /k+m)\gamma^{\nu}
(\rlap /k-\rlap /Q-m)\right]\over  (k-Q)^2-m^2+i\epsilon}\right]
\end{eqnarray}
for Majorana neutrinos, where in this latter case we put $n_{+}=n_{-}=n$ since the
chemical potential vanishes. Note that in (10) there is no component
proportional to $\epsilon^{\mu \nu \alpha \beta}$ since after the integration
over $k$ the only four-vector available is $Q^{\alpha}$. As a result there
will be no parity violating potentials.

Far from degeneracy (i.e. for chemical potential $\mu \ll T$), 
as is probably the case for cosmological
neutrinos, we can consider the neutrinos to be Boltzmann distributed, that is we take
\begin{equation}\label{boltz}
n_{\pm}=\exp[(\pm \mu -\vert k^0\vert)/T].
\end{equation}
With this approximation, the integrations involved in the calculation of potentials can
be easily done by conveniently choosing the order in which they are performed. The
results can be expressed again in terms of Bessel functions and are as follows:
\begin{equation}\label{diracT}
V_{T}^{Dirac}(r)=-{G_{F}^2m^4g_{V}g_{V}'\over \pi^3r}\cosh {(\mu/T)}
\left [{K_{1}(\rho)\over \rho}+{4K_{2}(\rho)\over \rho^2} \right]
\end{equation}
and
\begin{equation}\label{majoT}
V_{T}^{Majorana}(r)=-{4G_{F}^2m^4g_{V}g_{V}'\over \pi^3r}{K_{2}(\rho)\over \rho^2}
\end {equation}
where we have defined 
\begin{equation}
\rho\equiv {m\over T}\sqrt{1+(2rT)^2}.
\end{equation}

For massless neutrinos (and $\mu=0$) both potentials collapse to
\begin{equation}\label{zeromT}
V_{T}(r)=-{8G_{F}^2m^4g_{V}g_{V}'\over \pi^3r}{1\over \rho^4}
\end {equation}
which is the result given in reference \cite{hp}.
 Because the neutrino background
temperature is $T\sim (1.2 mm)^{-1}$, we see that for
distances much larger than $1mm$ (i.e. $rT\gg 1$) the potential in equation (16) reads
\begin{equation}\label{asympt}
V_{T}(r)\simeq -{G_{F}^2g_{V}g_{V}'\over 2\pi^3r^5}.
\end{equation}
When added to the vacuum result (9), the total potential is
\begin{equation}\label{totalV}
V_{tot}(r)\simeq -{G_{F}^2g_{V}g_{V}'\over 4\pi^3r^5}
\end {equation} 
that is, in the presence of the cosmic neutrino background the original Feinberg-Sucher
force switches sign, i.e. a repulsive force turns into an attractive one. On the other
hand, well within the sub-millimeter domain ($rT\ll 1$), the temperature dependent
potential (16) behaves as follows
\begin{equation}\label{submm}
V_{T}(r)\simeq -{8G_{F}^2g_{V}g_{V}'T^4\over \pi^3r}
\end{equation}
which is negligible compared to the vacuum contribution in equation (9). 

 In the general $m\ne 0$ case, we shall study the Dirac and Majorana potentials, equations
(13) and (14) respectively, in various physically interesting limits. Consider first the
cases where $r\ll m^{-1}\ll T^{-1} $ or $m^{-1}\ll r\ll T^{-1}$. Performing the relevant
expansions of the Bessel functions in (13) and (14) leads to
\begin{equation}\label{limD}
V_{T}^{Dirac}(r)\simeq -{G_{F}^2m^{5/2}g_{V}g_{V}'\over
  2^{1/2}\pi^{5/2}r}\;T^{3/2}\;
\cosh {(\mu/T)}e^{-m/T}
\end{equation}
and
\begin{equation}\label{limM}
V_{T}^{Majorana}(r)\simeq -{2^{3/2}G_{F}^2m^{3/2}g_{V}g_{V}'\over
  \pi^{5/2}r}\;T^{5/2}\;
e^{-m/T}.
\end{equation}
Hence, thermal effects are exponentially damped in both distance domains.

 A different behaviour is obtained for distances much larger than any inverse energy scale
in the problem, i.e. for $r\gg T^{-1} \gg m^{-1}$  or $r\gg m^{-1} \gg T^{-1}$. Indeed,
now we have
\begin{equation}\label{limD2}
V_{T}^{Dirac}(r)\simeq -{G_{F}^2g_{V}g_{V}'\over 4}\left( {m \over \pi r}
\right)^{5/2}
\cosh {(\mu/T)}e^{-2mr}
\end{equation}
and
\begin {equation}\label{limM2}
V_{T}^{Majorana}(r)\simeq -{G_{F}^2g_{V}g_{V}'\over2}\left({m^3\over
\pi^5r^7}\right)^{1/2}e^{-2mr}.
\end {equation}
Both expressions exhibit the characteristic Yukawa exponential damping associated to
two-particle exchange. These results when added to their vacuum counterparts, equations
(7) and (8), produce the inversion phenomenon already noticed in the massless case. At
asymptotically large distances the resulting potential is equal in strength as it would
be in vacuum but, contrary to what happens in vacuum, it is attractive instead.

 There is no exponential suppression only when $r\ll T^{-1}\ll m^{-1}$ or $T^{-1}\ll r \ll
m^{-1}$, where one essentially  recovers the massless cases, equations (19) or (17),
respectively. Indeed, for Majorana neutrinos one gets these equations as they stand,
and for Dirac neutrinos both equations should be multiplied by the factor
$\cosh{(\mu/T)}$ for non-zero chemical potential.

 Let us note that the results given in equations (20) and (22) for the $m\ne 0$ Dirac
case  disagree with the corresponding results given in reference \cite{hp}. Indeed,
their formulae do not show the Boltzmann or Yukawa suppression factors that enter the
asymptotic expansions of the Bessel functions and which are bound to be there on physical
grounds. For the sake of an explicit comparison we provide the reader, in the appendix
at the end of the paper, with some details of the calculation.  

 Up to this point all calculations refer to spin-independent potentials, those that can
coherently add over macroscopic samples of unpolarized matter. Let us, for the sake of
completeness, consider briefly the question of potentials that depend on spin. Now we
should focus on the spatial indices of the tensor $I_{\mu\nu}$  appearing in the
scattering amplitude in equation (2). The Fourier transformation (1) is in this case
somewhat more involved than before because the amplitude  will depend on the
components of the 3-momentum transfer. Nevertheless they can be easily performed and
we get:
\begin{equation}\label{potsd}
V_{T}^{spin}(r)
=-{4G_{F}^2m^4g_{A}g_{A}'\over \pi^3r}\left[({\bf S} \cdot {\bf S}')\:F(r)+2{({\bf S} \cdot
{\bf r}) ({\bf S}' \cdot {\bf r})\over r^2}\;G(r)\right]\cosh(\mu/T)
\end{equation}
where
\begin{equation}
F(r)\equiv a{K_{1}(\rho) \over \rho }+2{K_{2}(\rho)\over \rho^2}-8m^2r^2{K_{3}(\rho)
\over 
\rho^3}
\end{equation}
and
\begin{equation}
G(r)\equiv 7{K_{2}(\rho) \over \rho^2}-4m^2r^2{K_{3}(\rho) \over \rho^3}
\end{equation}
with $a=1$ for Dirac neutrinos and $a=2$ for Majorana neutrinos and ${\bf S}^2=3/4$.
Of course, in the Majorana case we must put $\mu=0$.

 Both cases above, i.e. Dirac and Majorana, lead to the potential
\begin{equation}\label{masszerosd}
V_{T}^{spin}(r)
=-{16G_{F}^2g_{A}g_{A}'T^4 \over \pi^3(1+4r^2T^2)^3r}\left[({\bf S}
  \cdot  {\bf S}')(1-12r^2T^2)+{({\bf S} \cdot {\bf r})({\bf S}' \cdot {\bf r})\over r^2}
(7+12r^2T^2)\right]
\end{equation}
for $m=0$ and $\mu=0$.
This result, eq. (27), should be then added to the vacuum result 
\cite{fs}
\begin{equation} \label{final}
V^{spin}(r)={G_{F}^2 g_{V}g_{V}' \over 2 \pi^3 r^5}\left[
5{({\bf S}\cdot{\bf r})({\bf S}'\cdot{\bf r}) 
\over r^2}-3({\bf S}\cdot{\bf S}')
\right].
\end{equation}
 The various regimes explored before can be studied also for the
spin-dependent forces. The discussion involves the various asymptotic forms
of the same modified Bessel functions and will lead to the same exponential
damping whenever the temperature or the mass is the relevant energy
parameter. Since these forces will be even more difficult to detect
than the spin-independent ones, for they do not add up coherently in
bulk matter, we do not bother here to display the explicit form for the
different limits.

 We end this paper with a short summary. Double neutrino exchange mediates (extremely
feeble) long range forces. In vacuum these forces have been known (at least for
Dirac neutrinos) for quite some time. Recently, it has been realised that a neutrino
background will also induce long range interactions among bulk matter. The
results were given for Dirac neutrinos. We have extended the work of Horowitz
and Pantaleone \cite{hp} 
to include the case of Majorana neutrinos and, furthermore, we
have derived the exact form of the potentials in either case, i.e. Dirac and
Majorana, and explored physically relevant distance and energy scales. In so
doing we have found important discrepancies with previous work ($m\ne0$, Dirac
case). Since matter is embedded in the cosmic neutrino background,
a consequence of our analysis is that the forces are repulsive in the
sub-millimeter scale and attractive for distances well beyond $1mm$ for any kind of
neutrino (massless or not). In fact, on the small scale the vacuum result
(Feinberg and Sucher) dominates whereas on the larger scale the relic neutrino
background is responsible for the dominant effect. This means that by experimentally
detecting (admittedly a highly improbable event for laboratory experiments) 
such forces in both different regimes
one would, not only  establish these neutrino interactions, but one  would in addition
detect the relic neutrino background. Actually it is the neutrino background temperature
($T^{-1}\sim1.2mm$) which sets this $1mm$ distance scale. 
 Incidentally, the sub-millimeter scale has been subject recently of renewed theoretical
as well as experimental interest \cite{dim}. 
For an experimental point of view of the actual
possible detection of very weak long range forces we refer the reader to references \cite{adel}.

{\bf Acknowledgements}

Work partially supported by the CICYT Research Projects AEN95-0815 and
AEN95-0882. F.F. acknowledges the CIRIT for financial support. M.N. would like
to thank the Spanish Ministerio de Educaci\'on y Ciencia.

\subsection*{Appendix}
\setcounter{equation}{0}
\renewcommand{\theequation}{A.\arabic{equation}}

We give here some details of the calculation for the massive Dirac case. All
integrals used can be found in \cite{grad}. 
Starting with eq. (1) we have (in
the formulae below $Q \equiv  |{\bf Q}|$):
\begin{eqnarray}
V_T^{Dirac} (r) &=& \frac{i G_F^2 g_v g_v'}{\pi^2 r} 
\int_0^{\infty}{dQ \:Q \:I_{00}(Q)
  \sin (Q r)} \nonumber \\
&=& \frac{G_F^2 g_v g'_v}{2 \pi^4 r} \cosh \left( \mu/T \right)
  \int_0^{\infty}{ \frac{dk k^2}{\sqrt{k^2+m^2}} \exp \left(-\sqrt{k^2+m^2}/T
  \right)} \nonumber \\
&& \times \int_{-1}^1 {dz \left[ (2 k z)^2-2 m^2-4 k^2 \right]
  \int_0^{\infty}{dQ\; \frac{Q \sin (Q r)}{Q^2-(2 k z)^2}}}
\end{eqnarray}

Performing first the integration over $Q$, then over $z$, we obtain:
\begin{equation}
V_T^{Dirac} (r)=
\frac{G_F^2 g_v g_v'}{2 \pi^3 r^4} \cosh \left( \mu/T \right) \left[
  -(1+(m r)^2)\; I (m,r,T)+r\: \frac{d I (m,r,T)}{d r} \right]
\end{equation}
where
\begin{eqnarray}
I(m,r,T)&=&\int_0^{\infty}{\frac{dk k}{\sqrt{k^2+m^2}}\exp
  \left(-\sqrt{k^2+m^2}/T \right) \sin (2 k r)} \nonumber \\
&= & \frac{ 2 r T m}{\sqrt{1+(2 r t)^2}}\;\; K_1 \left(\frac{m}{T} \sqrt{1+(2 r
  T)^2}\: \right)
\end{eqnarray}

Inserting (A.3) in (A.2) gives eq.(13).

For the spin-dependent part we decompose
\begin{equation}
T ({\bf Q})= ({\bf S}\cdot{\bf Q})({\bf S}'\cdot{\bf Q})\; t_1 (Q)+ 
({\bf  S}\cdot  {\bf S}' )\; t_2 (Q)
\end{equation}
where the functions $t_1 (Q)$ and $t_2 (Q)$ depend only on $|{\bf Q}|$
and we perform first the angular $\hat{Q}$ integration in eq.(1). The rest of
the calculation goes along similar lines as in the spin-independent part above.


\vskip 3cm

\begin{thebibliography}{99}
\bibitem{fs}
G. Feinberg and J. Sucher, Phys. Rev. {\bf A166} (1968) 1638;
G. Feinberg, J. Sucher and C.-K. Au, Phys. Rep. {\bf 180} (1989) 83
\bibitem{hs}
S. D. H. Hsu and P. Sikivie, Phys. Rev. {\bf D49} (1994) 4951
\bibitem{fey}
R. P. Feynman, F. B. Morinigo and W. G. Wagner, {\it `Feynman Lectures 
on Gravitation'}, Addison-Wesley, Readings, MA 1995
\bibitem{gri1}
J. A. Grifols, E. Masso and R., Toldra, Phys. Lett. {\bf B389} (1996) 363
\bibitem{fisch}
E. Fischbach, Ann. Phys. (N.Y.) {\bf 247} (1996) 213
\bibitem{gri2}
J. A. Grifols and S. Tortosa, Phys. Lett. {\bf B328} (1994) 98;
see also F. Ferrer and J. A. Grifols, ` Long Range Forces from
Pseudoscalar Exchange', UAB-FT-447, hep-ph/9805477
\bibitem{aba}
A. Abada, M. B. Gavela and O. Pene, Phys. Lett. {\bf B387} (1996) 315;
A. Y. Smirnov and F. Vissani, hep-ph/9604443
\bibitem{hp}
C. J. Horowitz and J. Pantaleone, Phys. Lett. {\bf B319} (1993) 186
\bibitem{dim}
S. Dimopoulos, M. Dine, S. Raby and S. Thomas, Phys. Rev. Lett.
{\bf 76} (1998) 70; I. Antoniadis, S. Dimopouls and G. Dvali, Nucl. Phys.
{\bf B516} (1998) 70
\bibitem{adel}
E. G. Adelberger, B. R. Heckel, C. W. Stubbs and W. F. Rogers,
Annu. Rev. Nucl. Part. Sci. {\bf 41} (1991) 269; C. I. Sukenik,
M. G. Boshier, D. Cho, V. Sandoghdar and E. A. Hinds, Phys. Rev. Lett.
{\bf 70} (1993) 560; J. C. Price  in {\it International Symposium on
Experimantal Gravitational Physics}, ed. P. F. Michelson, H. Enke and
G. Pizzella, 
World Scientific, 1987
\bibitem{grad}
I. S. Gradshteyn and I. M. Ryzhik, {\it Table of Integrals, Series and
Products}, corrected and enlarged edition, Academic Press, INC 1980
\end{thebibliography}
\end{document}